\documentclass[
10pt,
twocolumn,
superscriptaddress,
 amsmath,amssymb,
 aps,
pra,
]{revtex4-2}

\usepackage{cancel, comment}

\usepackage{amsmath}
\usepackage{mathtools}
\usepackage{listings}
\usepackage{amsfonts}
\usepackage{physics}
\usepackage{dsfont}
\usepackage{bm}
\usepackage{stackengine} 
\usepackage{enumerate}
\usepackage[colorlinks=true,linkcolor=blue, citecolor=blue, urlcolor=blue, bookmarks]{hyperref}

\usepackage{arydshln}
\usepackage{graphicx}

\usepackage[caption=false]{subfig}
\usepackage{array}
\usepackage{arydshln} 
\usepackage{graphicx}
\usepackage{tabularx} 
\usepackage{float}
\usepackage[table, svgnames]{xcolor}
\usepackage{adjustbox}
\usepackage{placeins}

\usepackage{tikz}
\usetikzlibrary{arrows,backgrounds,positioning,fit,matrix,angles,quotes}
\usepackage{pgfplots}
\pgfplotsset{compat=1.18}
\usetikzlibrary{external}

\DeclareMathOperator*{\oprod}{\bigotimes} 
\newcommand{\myvec}[1]{\boldsymbol{#1}}

\newcommand{\be}{\begin{equation}}
\newcommand{\ee}{\end{equation}}

\begin{document}

\title{Translation symmetry restoration in integrable systems: the noninteracting case}

\author{Molly Gibbins}
\author{Adam Gammon-Smith}
\affiliation{School of Physics and Astronomy, University of Nottingham, Nottingham, NG7 2RD, United Kingdom}
\affiliation{Centre for the Mathematics and Theoretical Physics of Quantum Non-Equilibrium Systems, University of Nottingham, Nottingham, NG7 2RD, United Kingdom}
\author{Bruno Bertini}
\affiliation{School of Physics and Astronomy, University of Birmingham, Birmingham, B15 2TT, United Kingdom}

\date{\today}

\begin{abstract}

The study of symmetry restoration has recently emerged as a fruitful means to extract high-level information on the relaxation of quantum many-body systems. However, while the restoration of internal symmetries has been investigated intensively, that of spatial symmetries has hitherto only been considered in the context of random unitary circuits. Here we present a complementary study of translation symmetry restoration in integrable systems. In particular, we consider a one-dimensional chain of spinless, non-interacting fermions quenched from a $\nu>1$ shift invariant state, and follow the local restoration of one-site shift invariance using the Frobenius distance $\Delta F_A$ between the state on a subsystem and its symmetrised counterpart. Distinct from the case of random unitary circuits, where symmetry restoration occurs abruptly for times proportional to the subsystem size, we find that symmetry here is restored smoothly and over timescales of the order of the subsystem size squared. We also find that the so-called `quantum Mpemba effect' is readily observed. Most importantly, we show that - in contrast to the case of continuous internal symmetries - this discrete symmetry restoration is not qualitatively described by a quasiparticle picture for $\Delta F_A$, and therefore goes beyond the hydrodynamic description. Our results can be directly extended to higher dimensions. 
\end{abstract}

\maketitle

{\textit{Introduction.---}} Studying the dynamics of quantum many-body systems is often complicated by the fact that local observables generate specific artifacts that are hard to filter out. Recent research has shown that a convenient way to circumvent this problem is to study the restoration of the symmetries of the dynamics that are broken by the initial state, as measured via the distance between the state of a local subsystem and its symmetrised counterpart~\cite{ares2022entanglement}. This has revealed a surprising phenomenon connected with the equilibration process: sometimes the symmetry is restored faster when the initial state is less symmetric. This effect bears a conceptual similarity with the Mpemba effect~\cite{mpemba1969cool}, occurring when water that is warmer freezes faster, and has been accordingly dubbed the quantum Mpemba effect (QME)~\cite{ares2022entanglement, ares2025quantum}. Since its first observation, QME has been theoretically predicted to occur in many different classes of systems ranging from integrable (free and interacting)~\cite{ares2023lack,murciano2024entanglement,khor2024confinement,ferro2023nonequilibrium,capizzi2023entanglement,capizzi2024universal,bertini2024dynamics,rylands2024microscopic,chen2024renyi,fossati2024entanglement,caceffo2024entangled,ares2024entanglement,yamashika2024entanglement, ares2024quantum,chalas2024multiple} to strongly non-integrable~\cite{liu2024symmetry, turkeshi2024quantum, klobas2024translation, foligno2025nonequilibrium} systems and even observed in trapped ion quantum simulators~\cite{joshi2024observing}.

Despite the intense research on the subject, most of the studies conducted so far have considered the restoration of internal symmetries --- $U(1)$ symmetries in the vast majority of the cases. This restriction comes from a technical, rather than conceptual, reason: namely, these symmetries generate particularly simple charges that simplify the analysis. Studying the restoration of other kinds of symmetries, however, can drastically widen the scope of this line of research. For instance, spatial symmetries are arguably a more generic feature of quantum many-body systems than internal ones. The study of spatial symmetry restoration has been initiated, and so far only considered, in Ref.~\cite{klobas2024translation}, which focussed on a class of ergodic (quantum chaotic) quantum systems --- random unitary circuits --- and demonstrated the occurrence of QME for intermediate sizes. Here we complement this study by considering translation symmetry restoration in integrable systems under constant Hamiltonian evolution. For the sake of simplicity, we focus on a simple non-interacting case. 

More specifically, we consider an infinite (or periodic) one-dimensional chain of spinless fermions on the lattice, whose dynamics are described by the tight binding Hamiltonian
\begin{equation}
\label{eq:hamiltonian}
H_{\rm TB} = \sum_{j} \left(c^\dagger_jc_{j+1} + \text{h.c.}\right).
\end{equation}
Although this Hamiltonian is invariant under single-site translations --- $[H_{\rm TB},T]=0$, where $T$ is the translation operator $T c_j T^{-1} = c_{j+1}$ --- we consider initial states that are only invariant under $\nu$-site shifts. Specifically, we consider product states of the form $\ket{\Psi} = \otimes_j \ket{\Psi_{\nu}}$, where $\ket{\Psi_{\nu}}$ is a Gaussian state defined on $\nu$ sites. The dynamics of the tight-binding model from states of this kind were recently considered in Ref.~\cite{gibbins2024quench}, which studied lattices of arbitrary dimension $d\geq 1$. Our upcoming discussion can similarly be generalised to higher dimensions.

Following Ref.~\cite{klobas2024translation}, to study the breaking and local restoration of the translation symmetry in a subsystem $A$, we use the \textit{Frobenius distance} between the reduced density matrix $\rho_A(t)$ and its symmetrised counterpart. These matrices are defined as
\begin{equation}
    \bar{\rho}_{A}(t) = \frac{1}{\nu} \sum_{j=0}^{\nu-1} \rho^{(j)}_{A}(t),\quad   \rho^{(j)}_{A}(t)\equiv \tr_{\bar{A}} \left[T^j \rho(t) T^{-j}\right],
    \label{eq:symmmat}
\end{equation}
where $\rho(t)=\ketbra{\Psi_t}$ is the state of the system at time $t$. Consequently, the (normalised) Frobenius distance reads

\begin{equation}
    \Delta F_A (t) \equiv \frac{\norm{\rho_{A}(t) - \bar{\rho}_{A}(t)}}{\norm{\rho_{A}(t)}}\,,
    \label{eq:frobasymm}
\end{equation}
where $\norm{A} \equiv \sqrt{\tr(AA^\dagger)}$ is the Frobenius norm and $\rho_A = \tr_{\bar{A}}(\rho)$ is the reduced density matrix. As discussed in Ref.~\cite{klobas2024translation}, when studying the restoration of spatial symmetries, Eq.~\eqref{eq:frobasymm} is more convenient than the entanglement asymmetry introduced in Ref.~\cite{ares2022entanglement} because the latter is much harder to access analytically. In our case, however, the two quantities become equivalent at leading order in $t$ and $|A|= l$. Furthermore, and which will be explicitly justified in due course, we note that the following discussion on spatial symmetry breaking should be compatible with any measure of symmetry breaking $f_A(\rho, t)$ that satisfies both $f_A \geq 0 \; \forall \;t$ and $f_A = 0 \iff \bar{\rho}_A = \rho_A$.  ~\footnote[1]{See the Supplemental Material for: (i) the definition of the entanglement asymmetry, generalised from the simplified form of~\cite{ares2023lack} for spatial symmetry breaking (ii) the explicit forms of both the normalised Frobenius distance and the $n=2$ entanglement asymmetry for the symmetry projected state of Eq.~\eqref{eq:symmmat} and (iii) the $t\to\infty$ expansion of the quasiparticle solution providing the scaling form of Eq.~\eqref{eq:scalingfunction}}. 

\begin{figure}[t]
    \includegraphics[width=\columnwidth]{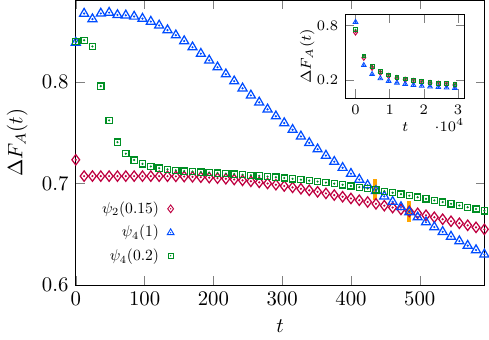}%
    \caption{The Frobenius distance $\Delta F_A(t)$, obtained via exact diagonalisation using Gaussian methods for subsystem size $l=100$ and different initial states. The states are of the form $\ket{\psi_{\nu}(\lambda)}$ where $\nu$ is the translational invariance and $\lambda$ is the superposition parameter, and their exact definitions are given in Eq.~\eqref{eq:states}. Crossings between the Frobenius distance of different states, which correspond to instances of QME, are indicated by the shaded orange boxes. Inset: Main plot extended to late times, showing that no further crossings occur.}
        \label{fig:frob}
\end{figure}

{\textit{Exact dynamics.---}}  The dynamics of $\Delta F_A (t)$ can be computed exactly by exploiting the Gaussianity of both the initial state and the dynamics, which allows us to write the Frobenius distance in terms of determinants of $l \times l$ matrices in the presence of number conservation~\cite{Note1}. For instance, in Fig.~\ref{fig:frob} we present the translation symmetry restoration in two families of states 
\be
\begin{aligned}
\ket{\psi_2 (\lambda_2)} &= \prod \frac{\left(c^\dag_{2 j} + \lambda_2 c^\dag_{1+2 j}\right)}{\sqrt{1+\lambda_2^2}} \ket{0},\\
\ket{\psi_4 (\lambda_4)} &= \prod c^\dag_{4 j} \frac{\left(c^\dag_{2+ 4 j} + \lambda_4 c^\dag_{1+ 4 j}\right)}{\sqrt{1+\lambda_4^2}} \ket{0}.
\end{aligned}
\label{eq:states}
\ee 

Interestingly, we see that the Frobenius distance decays as a power law in time in the asymptotic limit. This is in contrast with the sharp, initial-state-independent restoration at $t \propto l$ observed in random unitary circuits~\cite{klobas2024translation}. Moreover, the figure displays two distinct cases where QME is observed, as indicated by the shaded orange boxes. The later crossing occurs between an instance of $\ket{\psi_2 (\lambda_2)}$ and one of $\ket{\psi_4 (\lambda_4)}$, explicitly involving two states with different degrees of initial symmetry breaking. The earlier crossing occurs between two instances of $\ket{\psi_4 (\lambda_4)}$, and the amount of symmetry breaking is implicitly controlled by the superposition parameter $\lambda_4$. However, this figure also demonstrates that QME does not always occur, as it is absent between $\lambda_2=0.15$ and $\lambda_4 = 0.1$.

{\textit{Quasiparticle picture.---}} In many cases, the asymptotic dynamics of free-fermions at sufficiently large scales can be described by the quasiparticle picture \`a la Calabrese and Cardy~\cite{calabrese2005evolution}. The intuition behind this picture is that, in integrable systems, the large-scale dynamics can be described in terms of the motion of stable quasiparticles created by the initial state that move as free classical particles with fixed velocity. The correlations among the quasiparticles are generated by the initial state but, crucially, \emph{not by the dynamics}, which is able to reduce the calculation of how relevant quantities evolve in time to an exercise of simple kinematics ~\footnote[2]{Note that for Gaussian systems such as those under consideration here, the quasiparticle picture can be derived through an asymptotic expansion of the relevant determinants~\cite{fagotti2008evolution, caceffo2024entangled}.} Among other dynamical quantities, this picture is most celebrated for successfully capturing the entanglement growth~\cite{calabrese2005evolution,fagotti2008evolution, alba2017entanglement, bertini2022growth}, as well as the restoration of internal symmetries~\cite{ares2022entanglement, caceffo2024entangled}. Here we develop a similar approach to describe the asymptotic behaviour of $ \Delta F_A (t)$ in the case of translation symmetry restoration.

In our case, the quasiparticles are nothing but a semiclassical version of the momentum modes whose velocities are given by $v(k) = \epsilon'(k) = -2\sin(k)$, where $\epsilon(k)=\cos(k)$ is the dispersion relation of the Hamiltonian in Eq.~\eqref{eq:hamiltonian}. In order to assign them well-defined positions, one must operate at an `intermediate resolution' - dividing space into unit cells of size $\delta$ that satisfy $a \ll \delta \ll l$ for lattice spacing $a$ and performing a \textit{partial Fourier transform} of the initial state within each cell~\cite{bertini2018entanglement, gibbins2024quench}. The resulting picture then reduces to simple kinematics when the initial correlations are confined to small, regular groups of quasiparticles, as is the case for us: one finds that the translational symmetry of the initial state confers a $\nu$-plet structure to these correlations~\cite{gibbins2024quench}. More precisely, a given mode in the reduced Brillouin zone $k\in[0,2\pi/\nu)$ is correlated only with those from the same cell with $k' \in \{k+2\pi/\nu, k + 4\pi/\nu, \ldots, k + 2(\nu-1)\pi/\nu\}$.

Armed with this fact, we assume that the quench generates $\nu$-plets in localised in cells of size $\Delta \ll l$ and their full evolution can therefore be classically determined. This allows us to write the semiclassical representation of the time evolved state~\cite{bertini2018entanglement, gibbins2024quench} as $\rho_{A}(t) = \oprod_{x} \oprod_{p} {\rho}_A{(x, p)}\,$, where $x\in \mathbb Z_{L/\Delta}$, $p\in \frac{2\pi}{\Delta}\mathbb Z_{\Delta/\nu}$ and ${\rho}_A{(x, p)}$ is given in terms of the \textit{semiclassical modes} 
\begin{equation}
    c_{x, p}  = \frac{1}{\sqrt{\Delta}} \sum_{n \in  \mathbb Z^d_{\Delta} } e^{i  pn} c_{\Delta x+ n}, \quad p\in \frac{2\pi}{\Delta} \mathbb Z^d_{\Delta}\,. 
\label{eq:cellmodes}
\end{equation}

\begin{figure}[t]
    \includegraphics[width=\columnwidth]{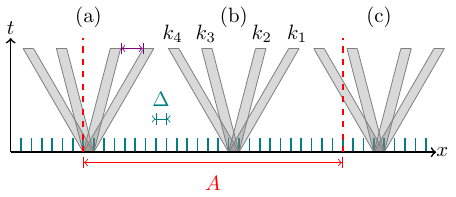}%
    \caption{Different bipartitions of a correlated multiplet for a $\nu=4$ initial state with respect to subsystem $A$. The multiplet is produced in a unit cell of size $\Delta$ and each mode $k_n(p) = p +2\pi n/\nu$ propagates with fixed velocity $v_n(p) = 2\sin(p +2\pi n/\nu)$. (a) A bipartition in which two modes are inside the subsystem. The purple dimension line indicates a \textit{relative velocity} between adjacent modes, which will be important for later discussion. (b) The bipartition in which all modes are inside the subsystem. This contributes to the Frobenius distance but not to the entanglement entropy (see main text). (c) A bipartition in which a single mode is inside the subsystem. This contributes to the entanglement entropy but not to the Frobenius distance.}
    \label{fig:bipartitions}
\end{figure}

More precisely, ${\rho}_A{(x, p)}$ is formed by considering each $\nu$-plet of correlated modes produced by the initial state tracing out all those that are out of the subsystem $A$ at time $t$~\cite{gibbins2024quench}. Note that, because $\ket{\Psi_0}$ is Gaussian, the correlations involving ${\rho}_A{(x, p)}$ can be computed from the two-point \textit{correlation matrix}

\begin{equation}
    [{C}_{A}(x,p)]_{k, k'} \equiv \tr\left( \rho^{\phantom{\dag}}_{A}(x,p) \hat{c}^{\dag}_{(x, p + k)} \hat{c}^{\phantom{\dag}}_{(x, p + k')} \right) \,. 
    \label{eq:corrmat}
\end{equation} 

To compute the Frobenius distance of Eq.~\eqref{eq:frobasymm}, we recast the $j$-site shifted operators $\rho^{(j)}_{A}(t)$ of ~Eq.~\eqref{eq:symmmat} in this semiclassical form, replacing ${\rho}_A{(x, p)}$ with ${\rho}^{(j)}_A(x, p) \equiv T_A^j {\rho}_A{(x, p)}T_A^{-j}$ where $T_A$ the periodic one-site translation operator in the subsystem $A$~\footnote[3]{Note that, at this level, a translation acts like an internal $\mathbb Z_\nu$ symmetry assigning charge $2 j\pi/\nu$ to each mode in $[2 j\pi/\nu,2 (j+1) \pi/\nu)$}. This eventually gives us~\cite{Note1} 
\begin{equation}
\Delta F_A (t)\big |_{\rm QP} = \sqrt{ \frac{(\nu-1)}{\nu} - \frac{1}{\nu} \sum_{j=1}^{\nu-1}r^{(j)}_A(t)},
\label{eq:quasiF}
\end{equation}
where we set
\begin{equation}
\!\!\!\!\! r^{(j)}_A(t) \!=\! \exp \left[\int\limits_{A\times[0,\frac{2\pi}{\nu}]}\!\!\!\!\!\!\!\!\frac{{{\rm d}x}{\rm d}k}{2\pi} \!\log\frac{\tr \left[{\rho}_A{(x, k)}{\rho}^{(j)}_A{(x, k)} \right]}{\tr \left[{\rho}^2_A{(x, k)} \right]}\! \right].
\label{eq:ratiofun}
\end{equation}
Two interesting comments are in order concerning this expression.

\begin{figure}[t]
    \centering
    \subfloat{%
        \includegraphics[width=\columnwidth]{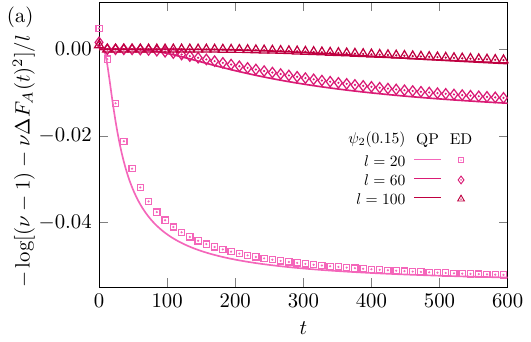}}\\[1ex]
    \caption{(a) Comparison of exact diagonalisation (data points) and quasiparticle method (solid lines) for $-\log[(\nu-1) - \nu \Delta F_A(t)^2]/l$ as subsystem size is increased, for $\psi_2(0.15)$. For (b) (the $\psi_4(1)$ state), refer to the Supplemental Material.}
    \label{fig:logratios}
\end{figure}

First, although Eqs.~\eqref{eq:quasiF} and \eqref{eq:ratiofun} allow one to compute the Frobenius distance by establishing how multiplets of correlated quasiparticles are split between $A$ and $\bar A$, the contributions controlling symmetry restoration differ from those controlling the saturation of spatial entanglement~\cite{calabrese2005evolution}. Indeed, while bipartitions where one single quasiparticle is either in or out of the system, e.g.~Fig.~\ref{fig:bipartitions} (a), contribute to the spreading of entanglement, they \textit{do not} contribute to Eqs.~\eqref{eq:quasiF} and \eqref{eq:ratiofun}. This is easily understood by viewing Eq.~\eqref{eq:ratiofun} together with Eq.~\eqref{eq:corrmat}: the translation operator is inert on a Gaussian single-particle density matrix, and so, for these bipartitions, the dynamical term in Eq.~\eqref{eq:quasiF} must vanish. Instead, rather than the longest-surviving bipartitions, constituting a single mode inside the subsystem, it is the next longest-surviving bipartitions --- those with exactly \textit{two} modes inside the subsystem, e.g. Fig.~\ref{fig:bipartitions} (c) ---   which determine the asymptotic behaviour of the Frobenius distance. On the other hand, it is longest-surviving bipartitions that characterise the hydrodynamic (fixed $t/l$) regime wherein the state of the subsystem is locally stationary~\cite{bertini2021finite, doyon2020lecture, bastianello2022introduction}. That these bipartitions do not enter into the symmetry restoration is a deceptively simple observation that nonetheless anticipates much of what will become apparent in this Letter: namely that --- even when studied \textit{within} the quasiparticle picture --- the symmetry restoration goes beyond hydrodynamics. It is also clear that statement is independent of the chosen measure of symmetry breaking, since any such measure can be written in terms of the moments of Eq.~\eqref{eq:ratiofun} for which the argument holds.

\begin{figure}[t]
    \centering
    \subfloat{%
        \includegraphics[width=\columnwidth]{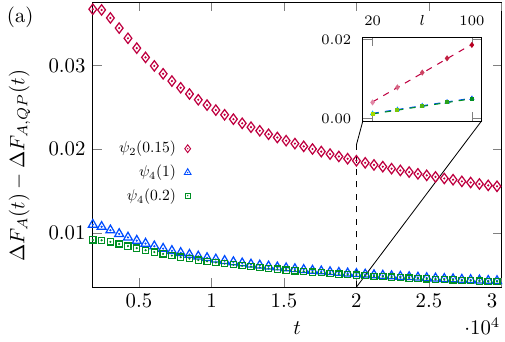}}\\[1ex]
    \subfloat{%
        \includegraphics[width=\columnwidth]{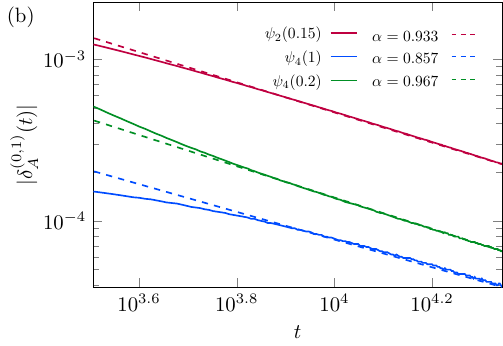}}
    \caption{(a) The difference between exact diagonalisation and quasiparticle solutions to the Frobenius distance, $\Delta F_A(t) -\Delta F_{A,\text{QP}}(t)$, for different initial states and subsystem size $l=100$. (b) The leading non-extensive correction $\delta^{(0,1)}_A(t)$, defined in Eq.~\eqref{eq:corrections}, for each of these inital states and subsystem size $l=200$. This panel is displayed on log-log scale at late times only, and the slopes of the dashed lines approximate the decay of the correction at large times as $t^{-\alpha}$. Inset (a): Example, shown for $t=20\,000$,  of the linear scaling of the difference $\Delta F_A(t) -\Delta F_{A,\text{QP}}(t)$ with system size.}
    \label{fig:deltafunc}
\end{figure}

\begin{figure}[t]
    \centering
    \subfloat{%
        \includegraphics[width=\columnwidth, trim=0 0 0 27, clip]{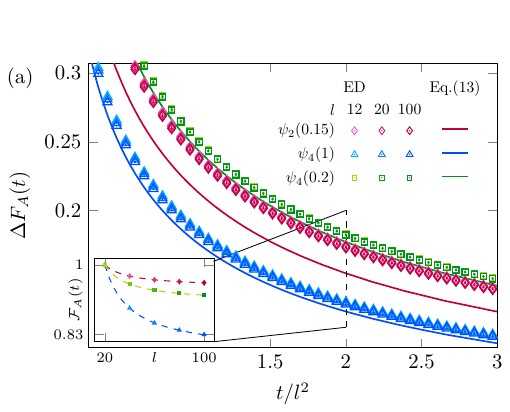}}\\[1ex]
    \subfloat{%
        \includegraphics[width=\columnwidth]{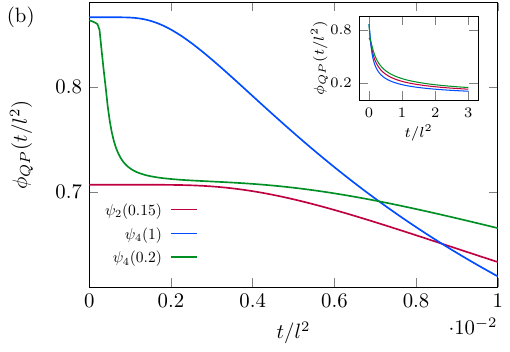}}
    \caption{The Frobenius distance in the scaling limit $t/l^2$ for different initial states and subsystem sizes. (a) compares the method of exact diagonalisation (data points) at increasing subsystem size to the functional form $\phi_{\rm QP}(\xi)$ given by Eq.~\eqref{eq:scalingfunction} obtained in the $l\to\infty$ limit of the quasiparticle method (solid lines), while (b) shows $\phi_{\rm QP}(\xi)$ at earlier times. Inset (a): Example, shown for $t/l^2=2$, of the scaling with system size compared to the prediction of $\phi_{\rm QP}(\xi)$, as defined via the normalised function $\mathcal{F}_A(t,l) = [\Delta F_A(t,l) - \phi_{\text{QP}}(\xi)] / [\Delta F_A(t,l_{min}) - \phi_{\text{QP}}(\xi)]$. The data are well described by a power-law fit of the form $al^{-b} + c$, where going from top to bottom we find $a=1.13, 2.02, 3.98$, $b=1.02, 0.97,1.03$, and $c=0.95,0.91, 0.78$. Inset (b): The main plot of (b) extended to late times, showing that no further crossings occur.}
        \label{fig:scaling}
\end{figure}

Second, Eq.~\eqref{eq:quasiF} involves two further approximations in addition to the basic quasiparticle picture: (i) In the definition of $\rho^{(j)}_{A}(x,k)$ we applied the translation only to the subsystem of interest. In other words, we assumed

\begin{equation}
\label{eq:enhancedQP}
\tr \left[\rho^{(i)}_{A}(t)\rho^{(j)}_{A}(t)\right] \simeq \tr \left[\rho_{A}(t) T_A^{j-i}\rho_{A}(t) T_A^{i-j}\right],  
\end{equation}

where $\simeq$ denotes equality at leading order in the subsystem size. This amounts to neglecting the boundary terms induced by a \textit{full-system} translation of the \textit{partial} Fourier creation operator. (ii) We computed the logarithm of the right-hand side\ of Eq.~\eqref{eq:enhancedQP} using the quasiparticle picture. Although one can rigorously show that this correctly accounts for the leading order in the subsystem size~\cite{fagotti2008evolution, caceffo2024entangled}, there are generically non-extensive corrections that we neglected. This step is inoffensive in the case of continuous symmetries --- see, e.g., Ref.~\cite{ares2023lack,rylands2024microscopic,caceffo2024entangled,ares2025quantum} ---where the leading contribution to $\Delta F_A (t)$ comes from a window of the integral over continuous charge sectors, analogous to the discrete summation in Eq.~\eqref{eq:quasiF}, in which the integration variable $\alpha$ is infinitesimal. In our case, however, the analogous summation only involves terms with finite $i-j$, and so forces us to reconsider this point.     

Concerning (i), we find that Eq.~\eqref{eq:enhancedQP} does indeed hold, with our results reported in Fig.~\ref{fig:logratios} demonstrating a rapid convergence of the shift-dependent quantity $-\log\left[(\nu-1)-\nu \Delta F_A(t)^2\right]/\ell$ obtained via exact diagonalisation (ED) to its quasiparticle picture (QP) prediction for increasing subsystem size. The innocent-looking neglect of non-extensive terms (ii), however, leads to substantial disagreements between ED and QP that appear to worsen for increasing values of $l$, as shown in the inset of Fig~\ref{fig:deltafunc}(a). This can be understood by noting that the corrections 

\begin{equation}
\label{eq:corrections}
\delta_{A}^{(i,j)}(t)=\log \frac{\tr\left[\rho^{(i)}_{A}(t)\rho^{(j)}_{A}(t)\right]}{\tr\left[\rho^{(0)}_{A}(t)^2\right]} - \log r^{(j-i)}_A(t), 
\end{equation}

generate non-trivial coefficients for the terms in the sum in Eq~\eqref{eq:quasiF}. In fact, as demonstrated by the second panel of Fig~\ref{fig:deltafunc}, for large times one has $\delta_{A}^{(i,j)}(t) \simeq t^{-\alpha}$ with $\alpha< 1$, while one can show that $\log r^{(j-i)}_A(t)  \simeq t^{-1}$~\cite{Note1}. This means that, for large enough times and any fixed $l$, the leading contribution to the ratio of traces is given by $\delta_{A}^{(i,j)}(t)$, not $\log r^{(j-i)}_A(t)$. Therefore, very surprisingly, we find that translation-symmetry restoration in non-interacting systems on the lattice, and the associated QME, are \emph{not} described by the quasiparticle picture.

{\textit{Scaling limit.---}} Although we saw that the quasiparticle picture does not describe the large time limit of a fixed subsystem, one can still wonder whether Eq~\eqref{eq:quasiF} provides a quantitative description of symmetry restoration in a certain `scaling limit'- i.e., when $t$ and $l$ are simultaneously taken to infinity along a particular curve. The usual `ballistic' scaling limit of $t,l\to\infty$ with fixed $t/l$ produces a trivial behaviour of the Frobenius distance, which remains pinned to the initial value and shows no symmetry restoration. Remarkably, however, Fig.~\ref{fig:scaling}(a) shows that the Frobenius distance approaches a well-defined scaling form in the `diffusive' scaling limit, where both $t$ and $l$ are simultaneously taken to infinity while keeping fixed $\xi=t/l^2$. This scaling limit explicitly emerges from the expansion of Eq.~\eqref{eq:ratiofun} in the $t\to\infty$ limit~\cite{Note1}, and provides an ideal setting for investigating the predictive power of the quasiparticle picture. In particular, noting that this limit takes the form
\begin{equation}
\label{eq:scalingfunction}
\lim_{l \to\infty} \log r^{(j)}_A(\xi l^2) = -\frac{a_j}{\xi}, \quad a_j\geq0,
\end{equation}

we see that $\phi_{\rm QP}(\xi) = \lim_{l \to\infty} \Delta F_A (\xi l^2)|_{\rm QP}$ is a positive function of $\xi$ that approaches 0 in the limit of infinite $\xi$, consistently describing a symmetry restoration process. This function bears explicit dependence on the initial state and, as shown in Fig.~\ref{fig:scaling} (b), is able to exhibit Mpemba-type crossings. 

To ascertain whether $\phi_{\rm QP}(\xi)$ gives an exact description of the symmetry restoration in the scaling limit, one should once again estimate the contribution of the corrections $\delta_{A}^{(i,j)}(t)$. Indeed, although the latter decay in time as $t^{-\alpha}$, they can in principle grow as $l^{\alpha+\beta}$, where $\beta<1$ is given by the sub-extensive nature of $\delta_{A}^{(i,j)}(t)$ in the ballistic scaling limit. Our numerical analysis suggests that $\beta=\alpha$ and, accordingly, the corrections give a finite contribution in the scaling limit: the inset of  Fig.~\ref{fig:scaling}(a) shows that the difference between the rescaled ED data and $\phi_{\rm QP}(\xi)$ does not seem to approach 0 for large $\ell$. Nevertheless, $\phi_{\rm QP}(\xi)$ is able to at least give a qualitative description of the symmetry restoration as it captures the correct timescale and remains close to the ED data for increasing $l$.

{\textit{Discussion.---}} Our results, which concern local restoration of translational symmetry in the tight-binding model, can be summarised as follows. First, we have seen that this symmetry restoration follows a polynomial law for large times --- in contrast with the exponential one observed for random unitary circuits~\cite{klobas2024translation} --- and can exhibit QME for arbitrary subsystem sizes. In fact, we observed that the Frobenius distance attains a non-trivial scaling form in the `diffusive' limit where time $t$ and subsystem size $l$ are simultaneously taken to infinity with fixed $\xi=t/l^2$. Second, we have found that --- although the system is trivially integrable  --- this symmetry restoration goes beyond the quasiparticle picture and is described by the latter only qualitatively in the diffusive scaling limit. This means in particular that for any given $l$ the asymptotic behaviour in time is \emph{not} described by the quasiparticle picture. We attributed this to the discrete nature of the symmetry that is restored. 

Although the quasiparticle description is only qualitative, it still gives insight into the physical mechanism behind the occurrence of QME in integrable systems. The multiplets of quasiparticles contributing to the Frobenius distance at late times are generically those with the smallest \textit{relative} velocities between modes. This matches the physical intuition: modes moving with very similar velocity spend more time close to each other and take longer to dephase~\cite{sotiriadis2016memory, sotiriadis2017equilibration}. The relaxation time is therefore determined by the density of modes with near-identical velocities, and by their contribution to the Frobenius distance. On the other hand, the initial value of the Frobenius distance depends on the full distribution of modes, thereby making it possible to manufacture initial states with greater initial Frobenius distance but shorter relaxation time, and vice versa, as required for QME. This mechanism generalises the one identified in Ref.~\cite{rylands2024microscopic} beyond the case of pair production and makes it obvious that QME is not just a property of the \emph{single particle} distribution of quasiparticles, but rather of the \emph{two-particle} one. This leads to the conclusion that QME is a phenomenon that fundamentally goes beyond the hydrodynamic description. 

We expect the phenomenology described here to be a generic feature of discrete translation symmetry restoration in integrable systems. The latter should also survive small integrability breaking terms provided that the quasiparticle lifetimes extend beyond the onset of the late-time regime (cf.~Ref.~\cite{bertini2020prethermalization}).

\begin{acknowledgments}
We thank Katja Klobas and Colin Rylands for useful discussions. We acknowledge financial support from the Royal Society through the University Research Fellowship No.\ 201101 (B.\ B.) and from UK Research and Innovation (UKRI) under the UK government's Horizon Europe funding guarantee [grant number EP/Y036069/1] (A.\ G.-S.). 
\end{acknowledgments}

\appendix

\bibliography{refs.bib}

\begin{thebibliography}{38}%
\makeatletter
\providecommand \@ifxundefined [1]{%
 \@ifx{#1\undefined}
}%
\providecommand \@ifnum [1]{%
 \ifnum #1\expandafter \@firstoftwo
 \else \expandafter \@secondoftwo
 \fi
}%
\providecommand \@ifx [1]{%
 \ifx #1\expandafter \@firstoftwo
 \else \expandafter \@secondoftwo
 \fi
}%
\providecommand \natexlab [1]{#1}%
\providecommand \enquote  [1]{``#1''}%
\providecommand \bibnamefont  [1]{#1}%
\providecommand \bibfnamefont [1]{#1}%
\providecommand \citenamefont [1]{#1}%
\providecommand \href@noop [0]{\@secondoftwo}%
\providecommand \href [0]{\begingroup \@sanitize@url \@href}%
\providecommand \@href[1]{\@@startlink{#1}\@@href}%
\providecommand \@@href[1]{\endgroup#1\@@endlink}%
\providecommand \@sanitize@url [0]{\catcode `\\12\catcode `\$12\catcode `\&12\catcode `\#12\catcode `\^12\catcode `\_12\catcode `\%12\relax}%
\providecommand \@@startlink[1]{}%
\providecommand \@@endlink[0]{}%
\providecommand \url  [0]{\begingroup\@sanitize@url \@url }%
\providecommand \@url [1]{\endgroup\@href {#1}{\urlprefix }}%
\providecommand \urlprefix  [0]{URL }%
\providecommand \Eprint [0]{\href }%
\providecommand \doibase [0]{https://doi.org/}%
\providecommand \selectlanguage [0]{\@gobble}%
\providecommand \bibinfo  [0]{\@secondoftwo}%
\providecommand \bibfield  [0]{\@secondoftwo}%
\providecommand \translation [1]{[#1]}%
\providecommand \BibitemOpen [0]{}%
\providecommand \bibitemStop [0]{}%
\providecommand \bibitemNoStop [0]{.\EOS\space}%
\providecommand \EOS [0]{\spacefactor3000\relax}%
\providecommand \BibitemShut  [1]{\csname bibitem#1\endcsname}%
\let\auto@bib@innerbib\@empty
\bibitem [{\citenamefont {Ares}\ \emph {et~al.}(2023{\natexlab{a}})\citenamefont {Ares}, \citenamefont {Murciano},\ and\ \citenamefont {Calabrese}}]{ares2022entanglement}%
  \BibitemOpen
  \bibfield  {author} {\bibinfo {author} {\bibfnamefont {F.}~\bibnamefont {Ares}}, \bibinfo {author} {\bibfnamefont {S.}~\bibnamefont {Murciano}},\ and\ \bibinfo {author} {\bibfnamefont {P.}~\bibnamefont {Calabrese}},\ }\bibfield  {title} {\bibinfo {title} {Entanglement asymmetry as a probe of symmetry breaking},\ }\href {https://doi.org/10.1038/s41467-023-37747-8} {\bibfield  {journal} {\bibinfo  {journal} {Nat. Commun.}\ }\textbf {\bibinfo {volume} {14}},\ \bibinfo {pages} {2036} (\bibinfo {year} {2023}{\natexlab{a}})}\BibitemShut {NoStop}%
\bibitem [{\citenamefont {Mpemba}\ and\ \citenamefont {Osborne}(1969)}]{mpemba1969cool}%
  \BibitemOpen
  \bibfield  {author} {\bibinfo {author} {\bibfnamefont {E.~B.}\ \bibnamefont {Mpemba}}\ and\ \bibinfo {author} {\bibfnamefont {D.~G.}\ \bibnamefont {Osborne}},\ }\bibfield  {title} {\bibinfo {title} {Cool?},\ }\href {https://doi.org/10.1088/0031-9120/4/3/312} {\bibfield  {journal} {\bibinfo  {journal} {Phys. Educ.}\ }\textbf {\bibinfo {volume} {4}},\ \bibinfo {pages} {172} (\bibinfo {year} {1969})}\BibitemShut {NoStop}%
\bibitem [{\citenamefont {Ares}\ \emph {et~al.}(2025{\natexlab{a}})\citenamefont {Ares}, \citenamefont {Calabrese},\ and\ \citenamefont {Murciano}}]{ares2025quantum}%
  \BibitemOpen
  \bibfield  {author} {\bibinfo {author} {\bibfnamefont {F.}~\bibnamefont {Ares}}, \bibinfo {author} {\bibfnamefont {P.}~\bibnamefont {Calabrese}},\ and\ \bibinfo {author} {\bibfnamefont {S.}~\bibnamefont {Murciano}},\ }\bibfield  {title} {\bibinfo {title} {The quantum mpemba effects},\ }\href {https://doi.org/10.1038/s42254-025-00838-0} {\bibfield  {journal} {\bibinfo  {journal} {Nature Reviews Physics}\ }\textbf {\bibinfo {volume} {7}},\ \bibinfo {pages} {451–460} (\bibinfo {year} {2025}{\natexlab{a}})}\BibitemShut {NoStop}%
\bibitem [{\citenamefont {Ares}\ \emph {et~al.}(2023{\natexlab{b}})\citenamefont {Ares}, \citenamefont {Murciano}, \citenamefont {Vernier},\ and\ \citenamefont {Calabrese}}]{ares2023lack}%
  \BibitemOpen
  \bibfield  {author} {\bibinfo {author} {\bibfnamefont {F.}~\bibnamefont {Ares}}, \bibinfo {author} {\bibfnamefont {S.}~\bibnamefont {Murciano}}, \bibinfo {author} {\bibfnamefont {E.}~\bibnamefont {Vernier}},\ and\ \bibinfo {author} {\bibfnamefont {P.}~\bibnamefont {Calabrese}},\ }\bibfield  {title} {\bibinfo {title} {Lack of symmetry restoration after a quantum quench: An entanglement asymmetry study},\ }\href {https://doi.org/10.21468/SciPostPhys.15.3.089} {\bibfield  {journal} {\bibinfo  {journal} {SciPost Phys.}\ }\textbf {\bibinfo {volume} {15}},\ \bibinfo {pages} {089} (\bibinfo {year} {2023}{\natexlab{b}})}\BibitemShut {NoStop}%
\bibitem [{\citenamefont {Murciano}\ \emph {et~al.}(2024)\citenamefont {Murciano}, \citenamefont {Ares}, \citenamefont {Klich},\ and\ \citenamefont {Calabrese}}]{murciano2024entanglement}%
  \BibitemOpen
  \bibfield  {author} {\bibinfo {author} {\bibfnamefont {S.}~\bibnamefont {Murciano}}, \bibinfo {author} {\bibfnamefont {F.}~\bibnamefont {Ares}}, \bibinfo {author} {\bibfnamefont {I.}~\bibnamefont {Klich}},\ and\ \bibinfo {author} {\bibfnamefont {P.}~\bibnamefont {Calabrese}},\ }\bibfield  {title} {\bibinfo {title} {Entanglement asymmetry and quantum {M}pemba effect in the {XY} spin chain},\ }\href {https://doi.org/10.1088/1742-5468/ad17b4} {\bibfield  {journal} {\bibinfo  {journal} {J. Stat. Mech.: Theory Exp.}\ }\textbf {\bibinfo {volume} {2024}}\bibinfo  {number} { (1)},\ \bibinfo {pages} {013103}}\BibitemShut {NoStop}%
\bibitem [{\citenamefont {Khor}\ \emph {et~al.}(2024)\citenamefont {Khor}, \citenamefont {K{\"u}rk{\c{c}}{\"u}oglu}, \citenamefont {Hobbs}, \citenamefont {Perdue},\ and\ \citenamefont {Klich}}]{khor2024confinement}%
  \BibitemOpen
\bibfield  {number} {  }\bibfield  {author} {\bibinfo {author} {\bibfnamefont {B.~J.}\ \bibnamefont {Khor}}, \bibinfo {author} {\bibfnamefont {D.}~\bibnamefont {K{\"u}rk{\c{c}}{\"u}oglu}}, \bibinfo {author} {\bibfnamefont {T.}~\bibnamefont {Hobbs}}, \bibinfo {author} {\bibfnamefont {G.}~\bibnamefont {Perdue}},\ and\ \bibinfo {author} {\bibfnamefont {I.}~\bibnamefont {Klich}},\ }\bibfield  {title} {\bibinfo {title} {Confinement and kink entanglement asymmetry on a quantum ising chain},\ }\href {https://doi.org/10.22331/q-2024-09-06-1462} {\bibfield  {journal} {\bibinfo  {journal} {Quantum}\ }\textbf {\bibinfo {volume} {8}},\ \bibinfo {pages} {1462} (\bibinfo {year} {2024})}\BibitemShut {NoStop}%
\bibitem [{\citenamefont {Ferro}\ \emph {et~al.}(2024)\citenamefont {Ferro}, \citenamefont {Ares},\ and\ \citenamefont {Calabrese}}]{ferro2023nonequilibrium}%
  \BibitemOpen
  \bibfield  {author} {\bibinfo {author} {\bibfnamefont {F.}~\bibnamefont {Ferro}}, \bibinfo {author} {\bibfnamefont {F.}~\bibnamefont {Ares}},\ and\ \bibinfo {author} {\bibfnamefont {P.}~\bibnamefont {Calabrese}},\ }\bibfield  {title} {\bibinfo {title} {Non-equilibrium entanglement asymmetry for discrete groups: The example of the {XY} spin chain},\ }\href {https://doi.org/10.1088/1742-5468/ad138f} {\bibfield  {journal} {\bibinfo  {journal} {J. Stat. Mech.: Theory Exp.}\ }\textbf {\bibinfo {volume} {2024}},\ \bibinfo {pages} {023101}}\BibitemShut {NoStop}%
\bibitem [{\citenamefont {Capizzi}\ and\ \citenamefont {Mazzoni}()}]{capizzi2023entanglement}%
  \BibitemOpen
  \bibfield  {author} {\bibinfo {author} {\bibfnamefont {L.}~\bibnamefont {Capizzi}}\ and\ \bibinfo {author} {\bibfnamefont {M.}~\bibnamefont {Mazzoni}},\ }\bibfield  {title} {\bibinfo {title} {{Entanglement asymmetry in the ordered phase of many-body systems: The {I}sing field theory}},\ }\href {https://doi.org/10.1007/JHEP12(2023)144} {\bibfield  {journal} {\bibinfo  {journal} {JHEP}\ }\textbf {\bibinfo {volume} {2023}},\ \bibinfo {pages} {144 (2023)}}\BibitemShut {NoStop}%
\bibitem [{\citenamefont {Capizzi}\ and\ \citenamefont {Vitale}(2024)}]{capizzi2024universal}%
  \BibitemOpen
  \bibfield  {author} {\bibinfo {author} {\bibfnamefont {L.}~\bibnamefont {Capizzi}}\ and\ \bibinfo {author} {\bibfnamefont {V.}~\bibnamefont {Vitale}},\ }\bibfield  {title} {\bibinfo {title} {A universal formula for the entanglement asymmetry of matrix product states},\ }\href {https://doi.org/10.1088/1751-8121/ad8796} {\bibfield  {journal} {\bibinfo  {journal} {J. Phys. A: Math. Theor.}\ }\textbf {\bibinfo {volume} {57}},\ \bibinfo {pages} {45LT01} (\bibinfo {year} {2024})}\BibitemShut {NoStop}%
\bibitem [{\citenamefont {Bertini}\ \emph {et~al.}(2024)\citenamefont {Bertini}, \citenamefont {Klobas}, \citenamefont {Collura}, \citenamefont {Calabrese},\ and\ \citenamefont {Rylands}}]{bertini2024dynamics}%
  \BibitemOpen
  \bibfield  {author} {\bibinfo {author} {\bibfnamefont {B.}~\bibnamefont {Bertini}}, \bibinfo {author} {\bibfnamefont {K.}~\bibnamefont {Klobas}}, \bibinfo {author} {\bibfnamefont {M.}~\bibnamefont {Collura}}, \bibinfo {author} {\bibfnamefont {P.}~\bibnamefont {Calabrese}},\ and\ \bibinfo {author} {\bibfnamefont {C.}~\bibnamefont {Rylands}},\ }\bibfield  {title} {\bibinfo {title} {Dynamics of charge fluctuations from asymmetric initial states},\ }\href {https://doi.org/10.1103/PhysRevB.109.184312} {\bibfield  {journal} {\bibinfo  {journal} {Phys. Rev. B}\ }\textbf {\bibinfo {volume} {109}},\ \bibinfo {pages} {184312} (\bibinfo {year} {2024})}\BibitemShut {NoStop}%
\bibitem [{\citenamefont {Rylands}\ \emph {et~al.}(2024)\citenamefont {Rylands}, \citenamefont {Klobas}, \citenamefont {Ares}, \citenamefont {Calabrese}, \citenamefont {Murciano},\ and\ \citenamefont {Bertini}}]{rylands2024microscopic}%
  \BibitemOpen
  \bibfield  {author} {\bibinfo {author} {\bibfnamefont {C.}~\bibnamefont {Rylands}}, \bibinfo {author} {\bibfnamefont {K.}~\bibnamefont {Klobas}}, \bibinfo {author} {\bibfnamefont {F.}~\bibnamefont {Ares}}, \bibinfo {author} {\bibfnamefont {P.}~\bibnamefont {Calabrese}}, \bibinfo {author} {\bibfnamefont {S.}~\bibnamefont {Murciano}},\ and\ \bibinfo {author} {\bibfnamefont {B.}~\bibnamefont {Bertini}},\ }\bibfield  {title} {\bibinfo {title} {Microscopic origin of the quantum {M}pemba effect in integrable systems},\ }\href {https://doi.org/10.1103/PhysRevLett.133.010401} {\bibfield  {journal} {\bibinfo  {journal} {Phys. Rev. Lett.}\ }\textbf {\bibinfo {volume} {133}},\ \bibinfo {pages} {010401} (\bibinfo {year} {2024})}\BibitemShut {NoStop}%
\bibitem [{\citenamefont {Chen}\ and\ \citenamefont {Chen}(2024)}]{chen2024renyi}%
  \BibitemOpen
  \bibfield  {author} {\bibinfo {author} {\bibfnamefont {M.}~\bibnamefont {Chen}}\ and\ \bibinfo {author} {\bibfnamefont {H.-H.}\ \bibnamefont {Chen}},\ }\bibfield  {title} {\bibinfo {title} {R\'enyi entanglement asymmetry in ($1+1$)-dimensional conformal field theories},\ }\href {https://doi.org/10.1103/PhysRevD.109.065009} {\bibfield  {journal} {\bibinfo  {journal} {Phys. Rev. D}\ }\textbf {\bibinfo {volume} {109}},\ \bibinfo {pages} {065009} (\bibinfo {year} {2024})}\BibitemShut {NoStop}%
\bibitem [{\citenamefont {Fossati}\ \emph {et~al.}(2024)\citenamefont {Fossati}, \citenamefont {Ares}, \citenamefont {Dubail},\ and\ \citenamefont {Calabrese}}]{fossati2024entanglement}%
  \BibitemOpen
  \bibfield  {author} {\bibinfo {author} {\bibfnamefont {M.}~\bibnamefont {Fossati}}, \bibinfo {author} {\bibfnamefont {F.}~\bibnamefont {Ares}}, \bibinfo {author} {\bibfnamefont {J.}~\bibnamefont {Dubail}},\ and\ \bibinfo {author} {\bibfnamefont {P.}~\bibnamefont {Calabrese}},\ }\bibfield  {title} {\bibinfo {title} {Entanglement asymmetry in {CFT} and its relation to non-topological defects},\ }\href {https://doi.org/10.1007/JHEP05(2024)059} {\bibfield  {journal} {\bibinfo  {journal} {J. High Energy Phys.}\ }\textbf {\bibinfo {volume} {2024}},\ \bibinfo {pages} {59}}\BibitemShut {NoStop}%
\bibitem [{\citenamefont {Caceffo}\ \emph {et~al.}(2024)\citenamefont {Caceffo}, \citenamefont {Murciano},\ and\ \citenamefont {Alba}}]{caceffo2024entangled}%
  \BibitemOpen
  \bibfield  {author} {\bibinfo {author} {\bibfnamefont {F.}~\bibnamefont {Caceffo}}, \bibinfo {author} {\bibfnamefont {S.}~\bibnamefont {Murciano}},\ and\ \bibinfo {author} {\bibfnamefont {V.}~\bibnamefont {Alba}},\ }\bibfield  {title} {\bibinfo {title} {Entangled multiplets, asymmetry, and quantum {M}pemba effect in dissipative systems},\ }\href {https://doi.org/10.1088/1742-5468/ad4537} {\bibfield  {journal} {\bibinfo  {journal} {J. Stat. Mech.: Theory Exp.}\ }\textbf {\bibinfo {volume} {2024}}\bibinfo  {number} { (6)},\ \bibinfo {pages} {063103}}\BibitemShut {NoStop}%
\bibitem [{\citenamefont {Ares}\ \emph {et~al.}(2024)\citenamefont {Ares}, \citenamefont {Murciano}, \citenamefont {Piroli},\ and\ \citenamefont {Calabrese}}]{ares2024entanglement}%
  \BibitemOpen
\bibfield  {number} {  }\bibfield  {author} {\bibinfo {author} {\bibfnamefont {F.}~\bibnamefont {Ares}}, \bibinfo {author} {\bibfnamefont {S.}~\bibnamefont {Murciano}}, \bibinfo {author} {\bibfnamefont {L.}~\bibnamefont {Piroli}},\ and\ \bibinfo {author} {\bibfnamefont {P.}~\bibnamefont {Calabrese}},\ }\bibfield  {title} {\bibinfo {title} {Entanglement asymmetry study of black hole radiation},\ }\href {https://doi.org/10.1103/PhysRevD.110.L061901} {\bibfield  {journal} {\bibinfo  {journal} {Phys. Rev. D}\ }\textbf {\bibinfo {volume} {110}},\ \bibinfo {pages} {L061901} (\bibinfo {year} {2024})}\BibitemShut {NoStop}%
\bibitem [{\citenamefont {Yamashika}\ \emph {et~al.}(2024)\citenamefont {Yamashika}, \citenamefont {Ares},\ and\ \citenamefont {Calabrese}}]{yamashika2024entanglement}%
  \BibitemOpen
  \bibfield  {author} {\bibinfo {author} {\bibfnamefont {S.}~\bibnamefont {Yamashika}}, \bibinfo {author} {\bibfnamefont {F.}~\bibnamefont {Ares}},\ and\ \bibinfo {author} {\bibfnamefont {P.}~\bibnamefont {Calabrese}},\ }\bibfield  {title} {\bibinfo {title} {Entanglement asymmetry and quantum {M}pemba effect in two-dimensional free-fermion systems},\ }\href {https://doi.org/10.1103/PhysRevB.110.085126} {\bibfield  {journal} {\bibinfo  {journal} {Phys. Rev. B}\ }\textbf {\bibinfo {volume} {110}},\ \bibinfo {pages} {085126} (\bibinfo {year} {2024})}\BibitemShut {NoStop}%
\bibitem [{\citenamefont {Ares}\ \emph {et~al.}(2025{\natexlab{b}})\citenamefont {Ares}, \citenamefont {Vitale},\ and\ \citenamefont {Murciano}}]{ares2024quantum}%
  \BibitemOpen
  \bibfield  {author} {\bibinfo {author} {\bibfnamefont {F.}~\bibnamefont {Ares}}, \bibinfo {author} {\bibfnamefont {V.}~\bibnamefont {Vitale}},\ and\ \bibinfo {author} {\bibfnamefont {S.}~\bibnamefont {Murciano}},\ }\bibfield  {title} {\bibinfo {title} {Quantum mpemba effect in free-fermionic mixed states},\ }\bibfield  {journal} {\bibinfo  {journal} {Physical Review B}\ }\textbf {\bibinfo {volume} {111}},\ \href {https://doi.org/10.1103/physrevb.111.104312} {10.1103/physrevb.111.104312} (\bibinfo {year} {2025}{\natexlab{b}})\BibitemShut {NoStop}%
\bibitem [{\citenamefont {Chalas}\ \emph {et~al.}(2024)\citenamefont {Chalas}, \citenamefont {Ares}, \citenamefont {Rylands},\ and\ \citenamefont {Calabrese}}]{chalas2024multiple}%
  \BibitemOpen
  \bibfield  {author} {\bibinfo {author} {\bibfnamefont {K.}~\bibnamefont {Chalas}}, \bibinfo {author} {\bibfnamefont {F.}~\bibnamefont {Ares}}, \bibinfo {author} {\bibfnamefont {C.}~\bibnamefont {Rylands}},\ and\ \bibinfo {author} {\bibfnamefont {P.}~\bibnamefont {Calabrese}},\ }\bibfield  {title} {\bibinfo {title} {Multiple crossings during dynamical symmetry restoration and implications for the quantum mpemba effect},\ }\href {https://doi.org/10.1088/1742-5468/ad769c} {\bibfield  {journal} {\bibinfo  {journal} {J. Stat. Mech.: Theory Exp.}\ }\textbf {\bibinfo {volume} {2024}}\bibinfo  {number} { (10)},\ \bibinfo {pages} {103101}}\BibitemShut {NoStop}%
\bibitem [{\citenamefont {Liu}\ \emph {et~al.}(2024)\citenamefont {Liu}, \citenamefont {Zhang}, \citenamefont {Yin},\ and\ \citenamefont {Zhang}}]{liu2024symmetry}%
  \BibitemOpen
\bibfield  {number} {  }\bibfield  {author} {\bibinfo {author} {\bibfnamefont {S.}~\bibnamefont {Liu}}, \bibinfo {author} {\bibfnamefont {H.-K.}\ \bibnamefont {Zhang}}, \bibinfo {author} {\bibfnamefont {S.}~\bibnamefont {Yin}},\ and\ \bibinfo {author} {\bibfnamefont {S.-X.}\ \bibnamefont {Zhang}},\ }\bibfield  {title} {\bibinfo {title} {Symmetry restoration and quantum {M}pemba effect in symmetric random circuits},\ }\href {https://doi.org/10.1103/PhysRevLett.133.140405} {\bibfield  {journal} {\bibinfo  {journal} {Phys. Rev. Lett.}\ }\textbf {\bibinfo {volume} {133}},\ \bibinfo {pages} {140405} (\bibinfo {year} {2024})}\BibitemShut {NoStop}%
\bibitem [{\citenamefont {Turkeshi}\ \emph {et~al.}(2025)\citenamefont {Turkeshi}, \citenamefont {Calabrese},\ and\ \citenamefont {De~Luca}}]{turkeshi2024quantum}%
  \BibitemOpen
  \bibfield  {author} {\bibinfo {author} {\bibfnamefont {X.}~\bibnamefont {Turkeshi}}, \bibinfo {author} {\bibfnamefont {P.}~\bibnamefont {Calabrese}},\ and\ \bibinfo {author} {\bibfnamefont {A.}~\bibnamefont {De~Luca}},\ }\bibfield  {title} {\bibinfo {title} {Quantum mpemba effect in random circuits},\ }\href {https://doi.org/10.1103/5d6p-8d1b} {\bibfield  {journal} {\bibinfo  {journal} {Phys. Rev. Lett.}\ }\textbf {\bibinfo {volume} {135}},\ \bibinfo {pages} {040403} (\bibinfo {year} {2025})}\BibitemShut {NoStop}%
\bibitem [{\citenamefont {Klobas}\ \emph {et~al.}(2025)\citenamefont {Klobas}, \citenamefont {Rylands},\ and\ \citenamefont {Bertini}}]{klobas2024translation}%
  \BibitemOpen
  \bibfield  {author} {\bibinfo {author} {\bibfnamefont {K.}~\bibnamefont {Klobas}}, \bibinfo {author} {\bibfnamefont {C.}~\bibnamefont {Rylands}},\ and\ \bibinfo {author} {\bibfnamefont {B.}~\bibnamefont {Bertini}},\ }\bibfield  {title} {\bibinfo {title} {Translation symmetry restoration under random unitary dynamics},\ }\href {https://doi.org/10.1103/PhysRevB.111.L140304} {\bibfield  {journal} {\bibinfo  {journal} {Phys. Rev. B}\ }\textbf {\bibinfo {volume} {111}},\ \bibinfo {pages} {L140304} (\bibinfo {year} {2025})}\BibitemShut {NoStop}%
\bibitem [{\citenamefont {Foligno}\ \emph {et~al.}(2025)\citenamefont {Foligno}, \citenamefont {Calabrese},\ and\ \citenamefont {Bertini}}]{foligno2025nonequilibrium}%
  \BibitemOpen
  \bibfield  {author} {\bibinfo {author} {\bibfnamefont {A.}~\bibnamefont {Foligno}}, \bibinfo {author} {\bibfnamefont {P.}~\bibnamefont {Calabrese}},\ and\ \bibinfo {author} {\bibfnamefont {B.}~\bibnamefont {Bertini}},\ }\bibfield  {title} {\bibinfo {title} {Nonequilibrium dynamics of charged dual-unitary circuits},\ }\href {https://doi.org/10.1103/PRXQuantum.6.010324} {\bibfield  {journal} {\bibinfo  {journal} {PRX Quantum}\ }\textbf {\bibinfo {volume} {6}},\ \bibinfo {pages} {010324} (\bibinfo {year} {2025})}\BibitemShut {NoStop}%
\bibitem [{\citenamefont {Joshi}\ \emph {et~al.}(2024)\citenamefont {Joshi}, \citenamefont {Franke}, \citenamefont {Rath}, \citenamefont {Ares}, \citenamefont {Murciano}, \citenamefont {Kranzl}, \citenamefont {Blatt}, \citenamefont {Zoller}, \citenamefont {Vermersch}, \citenamefont {Calabrese}, \citenamefont {Roos},\ and\ \citenamefont {Joshi}}]{joshi2024observing}%
  \BibitemOpen
  \bibfield  {author} {\bibinfo {author} {\bibfnamefont {L.~K.}\ \bibnamefont {Joshi}}, \bibinfo {author} {\bibfnamefont {J.}~\bibnamefont {Franke}}, \bibinfo {author} {\bibfnamefont {A.}~\bibnamefont {Rath}}, \bibinfo {author} {\bibfnamefont {F.}~\bibnamefont {Ares}}, \bibinfo {author} {\bibfnamefont {S.}~\bibnamefont {Murciano}}, \bibinfo {author} {\bibfnamefont {F.}~\bibnamefont {Kranzl}}, \bibinfo {author} {\bibfnamefont {R.}~\bibnamefont {Blatt}}, \bibinfo {author} {\bibfnamefont {P.}~\bibnamefont {Zoller}}, \bibinfo {author} {\bibfnamefont {B.}~\bibnamefont {Vermersch}}, \bibinfo {author} {\bibfnamefont {P.}~\bibnamefont {Calabrese}}, \bibinfo {author} {\bibfnamefont {C.~F.}\ \bibnamefont {Roos}},\ and\ \bibinfo {author} {\bibfnamefont {M.~K.}\ \bibnamefont {Joshi}},\ }\bibfield  {title} {\bibinfo {title} {Observing the quantum {M}pemba effect in quantum simulations},\ }\href {https://doi.org/10.1103/PhysRevLett.133.010402} {\bibfield  {journal} {\bibinfo  {journal} {Phys. Rev. Lett.}\ }\textbf
  {\bibinfo {volume} {133}},\ \bibinfo {pages} {010402} (\bibinfo {year} {2024})}\BibitemShut {NoStop}%
\bibitem [{\citenamefont {Gibbins}\ \emph {et~al.}(2024)\citenamefont {Gibbins}, \citenamefont {Jafarizadeh}, \citenamefont {Gammon-Smith},\ and\ \citenamefont {Bertini}}]{gibbins2024quench}%
  \BibitemOpen
  \bibfield  {author} {\bibinfo {author} {\bibfnamefont {M.}~\bibnamefont {Gibbins}}, \bibinfo {author} {\bibfnamefont {A.}~\bibnamefont {Jafarizadeh}}, \bibinfo {author} {\bibfnamefont {A.}~\bibnamefont {Gammon-Smith}},\ and\ \bibinfo {author} {\bibfnamefont {B.}~\bibnamefont {Bertini}},\ }\bibfield  {title} {\bibinfo {title} {Quench dynamics in lattices above one dimension: The free fermionic case},\ }\href {https://doi.org/10.1103/PhysRevB.109.224310} {\bibfield  {journal} {\bibinfo  {journal} {Phys. Rev. B}\ }\textbf {\bibinfo {volume} {109}},\ \bibinfo {pages} {224310} (\bibinfo {year} {2024})}\BibitemShut {NoStop}%
\bibitem [{Note1()}]{Note1}%
  \BibitemOpen
  \bibinfo {note} {See the Supplemental Material for: (i) the definition of the entanglement asymmetry, generalised from the simplified form of~\cite {ares2023lack} for spatial symmetry breaking (ii) the explicit forms of both the normalised Frobenius distance and the $n=2$ entanglement asymmetry for the symmetry projected state of Eq.~\protect \eqref {eq:symmmat} and (iii) the $t\to \infty $ expansion of the quasiparticle solution providing the scaling form of Eq.~\protect \eqref {eq:scalingfunction}}\BibitemShut {NoStop}%
\bibitem [{\citenamefont {Calabrese}\ and\ \citenamefont {Cardy}(2005)}]{calabrese2005evolution}%
  \BibitemOpen
  \bibfield  {author} {\bibinfo {author} {\bibfnamefont {P.}~\bibnamefont {Calabrese}}\ and\ \bibinfo {author} {\bibfnamefont {J.}~\bibnamefont {Cardy}},\ }\bibfield  {title} {\bibinfo {title} {Evolution of entanglement entropy in one-dimensional systems},\ }\href {https://doi.org/10.1088/1742-5468/2005/04/P04010} {\bibfield  {journal} {\bibinfo  {journal} {Journal of Statistical Mechanics: Theory and Experiment}\ }\textbf {\bibinfo {volume} {2005}},\ \bibinfo {pages} {P04010} (\bibinfo {year} {2005})}\BibitemShut {NoStop}%
\bibitem [{Note2()}]{Note2}%
  \BibitemOpen
  \bibinfo {note} {Note that for Gaussian systems such as those under consideration here, the quasiparticle picture can be derived through an asymptotic expansion of the relevant determinants~\cite {fagotti2008evolution, caceffo2024entangled}.}\BibitemShut {Stop}%
\bibitem [{\citenamefont {Fagotti}\ and\ \citenamefont {Calabrese}(2008)}]{fagotti2008evolution}%
  \BibitemOpen
  \bibfield  {author} {\bibinfo {author} {\bibfnamefont {M.}~\bibnamefont {Fagotti}}\ and\ \bibinfo {author} {\bibfnamefont {P.}~\bibnamefont {Calabrese}},\ }\bibfield  {title} {\bibinfo {title} {Evolution of entanglement entropy following a quantum quench: Analytic results for the $xy$ chain in a transverse magnetic field},\ }\href {https://doi.org/10.1103/PhysRevA.78.010306} {\bibfield  {journal} {\bibinfo  {journal} {Phys. Rev. A}\ }\textbf {\bibinfo {volume} {78}},\ \bibinfo {pages} {010306} (\bibinfo {year} {2008})}\BibitemShut {NoStop}%
\bibitem [{\citenamefont {Alba}\ and\ \citenamefont {Calabrese}(2017)}]{alba2017entanglement}%
  \BibitemOpen
  \bibfield  {author} {\bibinfo {author} {\bibfnamefont {V.}~\bibnamefont {Alba}}\ and\ \bibinfo {author} {\bibfnamefont {P.}~\bibnamefont {Calabrese}},\ }\bibfield  {title} {\bibinfo {title} {Entanglement and thermodynamics after a quantum quench in integrable systems},\ }\href {https://doi.org/10.1073/pnas.1703516114} {\bibfield  {journal} {\bibinfo  {journal} {Proc. Natl. Acad. Sci. U.S.A.}\ }\textbf {\bibinfo {volume} {114}},\ \bibinfo {pages} {7947} (\bibinfo {year} {2017})}\BibitemShut {NoStop}%
\bibitem [{\citenamefont {Bertini}\ \emph {et~al.}(2022)\citenamefont {Bertini}, \citenamefont {Klobas}, \citenamefont {Alba}, \citenamefont {Lagnese},\ and\ \citenamefont {Calabrese}}]{bertini2022growth}%
  \BibitemOpen
  \bibfield  {author} {\bibinfo {author} {\bibfnamefont {B.}~\bibnamefont {Bertini}}, \bibinfo {author} {\bibfnamefont {K.}~\bibnamefont {Klobas}}, \bibinfo {author} {\bibfnamefont {V.}~\bibnamefont {Alba}}, \bibinfo {author} {\bibfnamefont {G.}~\bibnamefont {Lagnese}},\ and\ \bibinfo {author} {\bibfnamefont {P.}~\bibnamefont {Calabrese}},\ }\bibfield  {title} {\bibinfo {title} {Growth of {R\'enyi} entropies in interacting integrable models and the breakdown of the quasiparticle picture},\ }\href {https://doi.org/10.1103/PhysRevX.12.031016} {\bibfield  {journal} {\bibinfo  {journal} {Phys. Rev. X}\ }\textbf {\bibinfo {volume} {12}},\ \bibinfo {pages} {031016} (\bibinfo {year} {2022})}\BibitemShut {NoStop}%
\bibitem [{\citenamefont {Bertini}\ \emph {et~al.}(2018)\citenamefont {Bertini}, \citenamefont {Fagotti}, \citenamefont {Piroli},\ and\ \citenamefont {Calabrese}}]{bertini2018entanglement}%
  \BibitemOpen
  \bibfield  {author} {\bibinfo {author} {\bibfnamefont {B.}~\bibnamefont {Bertini}}, \bibinfo {author} {\bibfnamefont {M.}~\bibnamefont {Fagotti}}, \bibinfo {author} {\bibfnamefont {L.}~\bibnamefont {Piroli}},\ and\ \bibinfo {author} {\bibfnamefont {P.}~\bibnamefont {Calabrese}},\ }\bibfield  {title} {\bibinfo {title} {Entanglement evolution and generalised hydrodynamics: noninteracting systems},\ }\href {https://doi.org/10.1088/1751-8121/aad82e} {\bibfield  {journal} {\bibinfo  {journal} {Journal of Physics A: Mathematical and Theoretical}\ }\textbf {\bibinfo {volume} {51}},\ \bibinfo {pages} {39LT01} (\bibinfo {year} {2018})}\BibitemShut {NoStop}%
\bibitem [{Note3()}]{Note3}%
  \BibitemOpen
  \bibinfo {note} {Note that, at this level, a translation acts like an internal $\protect \mathbb Z_\nu $ symmetry assigning charge $2 j\pi /\nu $ to each mode in $[2 j\pi /\nu ,2 (j+1) \pi /\nu )$}\BibitemShut {NoStop}%
\bibitem [{\citenamefont {Bertini}\ \emph {et~al.}(2021)\citenamefont {Bertini}, \citenamefont {Heidrich-Meisner}, \citenamefont {Karrasch}, \citenamefont {Prosen}, \citenamefont {Steinigeweg},\ and\ \citenamefont {\ifmmode \check{Z}\else \v{Z}\fi{}nidari\ifmmode~\check{c}\else \v{c}\fi{}}}]{bertini2021finite}%
  \BibitemOpen
  \bibfield  {author} {\bibinfo {author} {\bibfnamefont {B.}~\bibnamefont {Bertini}}, \bibinfo {author} {\bibfnamefont {F.}~\bibnamefont {Heidrich-Meisner}}, \bibinfo {author} {\bibfnamefont {C.}~\bibnamefont {Karrasch}}, \bibinfo {author} {\bibfnamefont {T.}~\bibnamefont {Prosen}}, \bibinfo {author} {\bibfnamefont {R.}~\bibnamefont {Steinigeweg}},\ and\ \bibinfo {author} {\bibfnamefont {M.}~\bibnamefont {\ifmmode \check{Z}\else \v{Z}\fi{}nidari\ifmmode~\check{c}\else \v{c}\fi{}}},\ }\bibfield  {title} {\bibinfo {title} {Finite-temperature transport in one-dimensional quantum lattice models},\ }\href {https://doi.org/10.1103/RevModPhys.93.025003} {\bibfield  {journal} {\bibinfo  {journal} {Rev. Mod. Phys.}\ }\textbf {\bibinfo {volume} {93}},\ \bibinfo {pages} {025003} (\bibinfo {year} {2021})}\BibitemShut {NoStop}%
\bibitem [{\citenamefont {Doyon}(2020)}]{doyon2020lecture}%
  \BibitemOpen
  \bibfield  {author} {\bibinfo {author} {\bibfnamefont {B.}~\bibnamefont {Doyon}},\ }\bibfield  {title} {\bibinfo {title} {{Lecture notes on Generalised Hydrodynamics}},\ }\href {https://doi.org/10.21468/SciPostPhysLectNotes.18} {\bibfield  {journal} {\bibinfo  {journal} {SciPost Phys. Lect. Notes}\ ,\ \bibinfo {pages} {18}} (\bibinfo {year} {2020})}\BibitemShut {NoStop}%
\bibitem [{\citenamefont {Bastianello}\ \emph {et~al.}(2022)\citenamefont {Bastianello}, \citenamefont {Bertini}, \citenamefont {Doyon},\ and\ \citenamefont {Vasseur}}]{bastianello2022introduction}%
  \BibitemOpen
  \bibfield  {author} {\bibinfo {author} {\bibfnamefont {A.}~\bibnamefont {Bastianello}}, \bibinfo {author} {\bibfnamefont {B.}~\bibnamefont {Bertini}}, \bibinfo {author} {\bibfnamefont {B.}~\bibnamefont {Doyon}},\ and\ \bibinfo {author} {\bibfnamefont {R.}~\bibnamefont {Vasseur}},\ }\bibfield  {title} {\bibinfo {title} {Introduction to the special issue on emergent hydrodynamics in integrable many-body systems},\ }\href {https://doi.org/10.1088/1742-5468/ac3e6a} {\bibfield  {journal} {\bibinfo  {journal} {Journal of Statistical Mechanics: Theory and Experiment}\ }\textbf {\bibinfo {volume} {2022}},\ \bibinfo {pages} {014001} (\bibinfo {year} {2022})}\BibitemShut {NoStop}%
\bibitem [{\citenamefont {Sotiriadis}(2016)}]{sotiriadis2016memory}%
  \BibitemOpen
  \bibfield  {author} {\bibinfo {author} {\bibfnamefont {S.}~\bibnamefont {Sotiriadis}},\ }\bibfield  {title} {\bibinfo {title} {Memory-preserving equilibration after a quantum quench in a one-dimensional critical model},\ }\href {https://doi.org/10.1103/PhysRevA.94.031605} {\bibfield  {journal} {\bibinfo  {journal} {Phys. Rev. A}\ }\textbf {\bibinfo {volume} {94}},\ \bibinfo {pages} {031605} (\bibinfo {year} {2016})}\BibitemShut {NoStop}%
\bibitem [{\citenamefont {Sotiriadis}(2017)}]{sotiriadis2017equilibration}%
  \BibitemOpen
  \bibfield  {author} {\bibinfo {author} {\bibfnamefont {S.}~\bibnamefont {Sotiriadis}},\ }\bibfield  {title} {\bibinfo {title} {Equilibration in one-dimensional quantum hydrodynamic systems},\ }\href {https://doi.org/10.1088/1751-8121/aa8aa5} {\bibfield  {journal} {\bibinfo  {journal} {Journal of Physics A: Mathematical and Theoretical}\ }\textbf {\bibinfo {volume} {50}},\ \bibinfo {pages} {424004} (\bibinfo {year} {2017})}\BibitemShut {NoStop}%
\bibitem [{\citenamefont {Bertini}\ and\ \citenamefont {Calabrese}(2020)}]{bertini2020prethermalization}%
  \BibitemOpen
  \bibfield  {author} {\bibinfo {author} {\bibfnamefont {B.}~\bibnamefont {Bertini}}\ and\ \bibinfo {author} {\bibfnamefont {P.}~\bibnamefont {Calabrese}},\ }\bibfield  {title} {\bibinfo {title} {Prethermalization and thermalization in entanglement dynamics},\ }\href {https://doi.org/10.1103/PhysRevB.102.094303} {\bibfield  {journal} {\bibinfo  {journal} {Phys. Rev. B}\ }\textbf {\bibinfo {volume} {102}},\ \bibinfo {pages} {094303} (\bibinfo {year} {2020})}\BibitemShut {NoStop}%
\end{thebibliography}%

\onecolumngrid

\vspace{1cm}
\noindent\rule{\linewidth}{0.4pt}
\vspace{0.5cm}
\begin{center}
{\bf\large Supplemental Material:\ Translation symmetry restoration in integrable systems: the noninteracting case}
\end{center}

In this Supplemental Material, we report some results complementing the main text. In particular
\begin{itemize}
\item In Sec.~1 we present the plot of Fig.~\ref{fig:logratios}(b) for the $\psi_4(1)$ state, which, together with Fig.~\ref{fig:logratios}(a), demonstrates the rapid approach to zero of extensive corrections generated by the approximation of Eq.~\eqref{eq:enhancedQP}.
\item  In Sec.~2 we show the explicit calculation that relates our chosen measure, the normalised Frobenius distance $\Delta F_A(\rho,t)$ of Eq.~\eqref{eq:frobasymm}, to the two-point correlations of Eq.~\eqref{eq:corrmat}.
\item  In Sec.~3 we focus on the families of states $\ket{\psi_2(\lambda_2)}$ and $\ket{\psi_4(\lambda_4)}$ introduced in~Eq.~\eqref{eq:states} and determine the constraints on the superposition parameters $\lambda_2, \lambda_4$ in order for a crossing of the Frobenius distance, as approximated by the quasiparticle picture, to occur between these states.
\end{itemize}

\section{Extensive corrections for the $\psi_4(1)$ state}
\label{sm:fig2b}

\begin{figure}[h]
    \centering
    \begin{minipage}{0.5\columnwidth} 
        \centering
        \subfloat{\includegraphics[width=\linewidth]{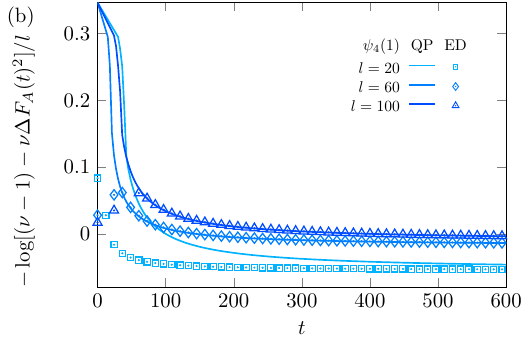}}
        \captionsetup{justification=raggedright, singlelinecheck=false, width=\linewidth}
        \renewcommand{\thefigure}{3(b)}
        \caption{Comparison of exact diagonalisation (data points) and quasiparticle method (solid lines)
        for $-\log[((\nu-1) - \nu \Delta F_A(t)^2)]/l$ as subsystem size is increased, for the state $\psi_4(1)$.}
    \end{minipage}
\end{figure}

\section{Frobenius Distance}
\label{sm:distance}

Applying the definition of the symmetrised density matrix $\bar{\rho}_A(t)$ from Eq.~\eqref{eq:symmmat} to Eq.~\eqref{eq:frobqp} yields

\begin{equation}
\Delta F_A(t)^2 =  1 + \frac{1}{\nu^2} \frac{\tr_A \left( \sum_{i,j=0}^{\nu - 1}\rho_A^{(i)}(t)\rho_A^{(j)}(t) \right)}{\tr_A (\rho_A(t)^2 )} - \frac{2}{\nu} \frac{\tr_A\left( \sum_{j=0}^{\nu - 1} \rho_A^{\vphantom{j}}(t) \rho_A^{(j)}(t) \right)}{ \tr_A (\rho_A(t)^2) } \, .
\label{eq:frobexpansion}
\end{equation}

Using that $\rho_A^{(j)}(t)$ is Gaussian for free-fermionic states, one can apply the composition rules for the trace of a product of Gaussian matrices \cite{ares2023lack} in order to rewrite the $\tr_A(\rho_A^{(i)}(t) \rho_A^{(j)}(t))$ as

\begin{equation}
    \tr_A(\rho_A^{(i)}(t) \rho_A^{(j)}(t)) = \det\left[(1-C_A^{(i)}(t))(1-C_A^{(j)}(t)) + C_A^{(i)}(t) C_A^{(j)}\!(t) \right] \, ,
    \label{eq:moments}
\end{equation}

where $[C_A^{(j)}(t)]_{n,m} \equiv \tr_A \left[c^{\dag}_n c^{\phantom{\dag}}_m \rho_A^{(j)}(t) \right]$ is the correlation matrix of subsystem $A$ under $j$ site shifts of the \textit{full} system.

\subsection{Quasiparticle Solution}
\label{sm:qpsol}

To proceed with the quasiparticle solution, anticipating the transform to the momentum modes of the system, we apply the approximation of Eq.~\eqref{eq:enhancedQP} to Eq.~\eqref{eq:frobexpansion} and arrive at the simplified expression

\begin{equation}
    \Delta F_A(t)^2 = \frac{1}{\nu} \left( (\nu-1) - \sum_{j=1}^{\nu-1}\frac{\tr_A(\rho_A^{\vphantom{j}}(t) \rho_A^{(j)}(t))}{\tr_{A}(\rho_A(t)^2)}\right) \, ,
    \label{eq:frobqp}
\end{equation}

which coincides with Eq.~\eqref{eq:quasiF} upon replacing $\rho_A^{\vphantom{j}}(t), \rho_A^{(j)}(t)$ with their semiclassical approximations given by Eq.~\eqref{eq:cellmodes}.  Namely, the quasiparticle picture of $\Delta F_A(t)$ is given by the set of $\nu$ shift-dependent trace terms

\begin{equation}
    Z^{(j)}_A(t) \equiv \tr_A(\rho_A^{(0)}(t) \rho_A^{(j)}(t))\, .
\label{eq:transmoments}
\end{equation}

We refer to Eq.~\eqref{eq:transmoments} as the \textit{translational moments} of the system, drawing a direct analogue to the so-called \textit{charged moments} of $U(1)$ symmetry breaking \cite{ares2022entanglement}. It is instructive to compare Eq.~\eqref{eq:frobqp} to the entanglement asymmetry $\Delta S^{(n)}_A(t)$ for $n=2$,

\begin{equation}
    \Delta S^{(n)}_A(t) \equiv \ln\tr_A\left(\frac{\rho_A^{n+1}}{n}\right) - \ln\tr_A\left(\frac{\rho_A^{\vphantom{n}}\bar{\rho}_A^{n}}{n}\right) \,\rightarrow
\end{equation}

\begin{equation}
    \Delta S^{(2)}_A(t) = \ln \left[\frac{1}{\nu} \left( 1 + \sum_{j=1}^{\nu-1} \frac{\tr_A(\rho_A^{\vphantom{j}}(t) \rho_A^{(j)}(t))}{\tr_A(\rho_A(t)^2)}\right) \right] \, ,
    \label{eq:renasymm}
\end{equation}

since it is apparent that, following the approximation of Eq.~\eqref{eq:enhancedQP}, these two measures are in fact equivalent at leading order in $t$ and $l$. As noted in \cite{klobas2024translation}, however, $\Delta S^{(2)}_A(t)$ becomes much harder to compute analytically whenever the simplification in \cite{ares2023lack} does not apply, making $\Delta F_A(t)$ our measure of choice.

Returning to the translational moments, these are computed in the quasiparticle picture by their semiclassical approximation

\begin{equation}
    Z^{(j)}_A(t) = \prod_{k,x,\myvec{n}} \tilde{Z}^{(j)}_{\myvec{n}}(k) \cdot \chi_{\myvec{n}} (A, k, x, t) \, ,
    \label{eq:qpdynamicalmoments}
\end{equation}

where $\tilde{Z}^{(j)}_{\myvec{n}}(k)$ is the moment associated with the cell mode bipartition $\myvec{n}$ under $j$ site shifts of the \textit{subsystem}

\begin{equation}
    \tilde{Z}^{(j)}_{\myvec{n}}(k) \equiv  \det\left[(1-\tilde{C}_{\myvec{n}}^{\vphantom{(j)}}(k))(1-\tilde{C}_{\myvec{n}}^{(j)}(k)) + \tilde{C}_{\myvec{n}}^{\vphantom{(j)}}(k) \tilde{C}_{\myvec{n}}^{(j)}\!(k) \right] \, ,
    \label{eq:kmoments}
\end{equation}

and $\chi_{\myvec{n}} (A, k, x, t)$ is the characteristic function of cell mode biparition $\myvec{n}$ with respect to the subsystem at time $t$

\begin{equation}
     \chi_{\myvec{n}} (k, x, t) = \prod_{\substack{n \in \myvec{n}\\ \bar{n} \notin \myvec{n}}} \chi_{A} (x_n(k,t)) \cdot \chi_{\bar{A}} (x_{\bar{n}}(k,t)) \, .
\end{equation}

Finally, using that the sum of this characteristic function over $x$ gives the areas $\mathcal{A}_{\myvec{n}} (A,k,t)$ associated with each cell mode bipartition, we see that the ratio function of Eq.~\eqref{eq:ratiofun} becomes 

\begin{equation}
    r^{(j)}_A(t) = \frac{Z^{(j)}(t)}{Z^{(0)}(t)} = \exp \left[\int_0^{\frac{2\pi}{\nu}}\frac{{\rm d}k}{2\pi} \mathcal{A}_{\myvec{n}}(A,k,t)\log \frac{\tilde{Z}^{(j)}_{\myvec{n}}(k)}{\tilde{Z}^{(0)}_{\myvec{n}}(k)} \right].
\end{equation}

Consequently, this solution contains all the features of a standard quasiparticle solution familiar to us from its application to the entanglement entropy \cite{calabrese2005evolution}, wherein each $\log(\tilde{Z}^{(j)}_{\myvec{n}}(k)/\tilde{Z}^{(0)}_{\myvec{n}}(k))$ is the contribution from each cell mode bipartition to the Frobenius distance, and the set of $\mathcal{A}_{\myvec{n}} (A,k,t)$ encodes the dynamical evolution of the cell modes with respect to subsystem $A$. Armed with these two ingredients, the integral over all modes $k$ thus provides a complete picture of the evolution of the Frobenius distance reconstructed with a single $\nu \times \nu$ matrix $\tilde{C}_{\myvec{n}}(k)$. 

\subsection{Numerical Solution}
\label{sm:edsol}

On the other hand, we can obtain the time-dependent correlations $C_A(t)$ of Eq.~\eqref{eq:moments} directly from exact diagonalisation. For purposes of generality, we consider a general superposition state

\begin{equation}
    \ket{\psi} = \left( \prod_{j=0}^{L/\nu - 1} \sum_{r=0}^{\nu-1} \alpha_r c^\dag_{\nu j - r}\right) \ket{0} \, ,
    \label{eq:initgeneral}
\end{equation}

where $\alpha_r$ are the set of $\nu$ superposition parameters within each unit cell and the state is normalised as $\sum_r \alpha_r^2 = 1$. The calculation is straightforward. The two-point correlations of this initial state in position space,

\begin{equation}
    \bra{\psi} c^\dag_n c_m \ket{\psi} = \sum_{j, r, s} \delta_{n, \nu j - r} \delta_{n, \nu j - s} \alpha_r \alpha_s^* \,,
\end{equation}

yield the two-point correlations in momentum space

\begin{equation}
    \bra{\psi} c^\dag_p c_{p'}^{\vphantom{\dag}} \ket{\psi} = \left( \frac{1}{\nu} \sum_{r,s = 0}^{\nu -1} e^{i(rp-sp')} \alpha_r \alpha_s^*\right) \left(\sum_{k \in \frac{2\pi}{\nu} \mathbb{Z}_\nu} \hspace{-5pt} \delta_{p, p'+k}\right) \,.
\end{equation}

This gives the time-dependent correlations, in the thermodynamic limit, as

\begin{equation}
    [C(t)]_{n,m} \equiv \lim_{L \to \infty} \bra{\psi} c^\dag_n (t) c_m^{\vphantom{\dag}} (t)\ket{\psi} = \frac{1}{\nu} \sum_{r,s = 0}^{\nu -1} \int_0^{2\pi} \frac{dp}{2\pi} \sum_{k \in \frac{2\pi}{\nu} \mathbb{Z}_\nu}  e^{i(n-m)p} e^{i(r-s)p} e^{i(m+s)k} e^{it[\epsilon(p-k)-\epsilon(p)]} \alpha_r \alpha_s^*
\end{equation}

and the matrix on the subsystem $C_A(t)$ is simply obtained from the restriction $n, m \in (0,1...l)$ where $\abs{A} \equiv l$. 

\section{Approximation of Frobenius Distance Crossings}

We choose to focus on the families of states $\ket{\psi_2(\lambda_2)}$ and $\ket{\psi_4(\lambda_4)}$, defined in Eq.~\eqref{eq:states}, as a means to observe QME between states of equal charge density and different symmetry breaking. In particular, $\lambda_4$  controls the $\nu=2$ symmetry breaking of the state, with $\lambda_2 = \lambda_4 = 0$ corresponding to the N\'eel state and thus providing a useful benchmark whereby we can expect agreement between the solutions of these two families of states.

\subsection{Initial Frobenius Distance}
\label{sm:init}

We begin with the states $\ket{\psi_2(\lambda_2)}$. The moments of Eq.~\eqref{eq:kmoments} are given for these $\nu=2$ states by the two-point correlators $n_k = \langle c^{\dag}_k c_k \rangle$ as

\begin{equation}
     \tilde{Z}^{(j)}_{\myvec{n}}(k) = \abs{\bra{\psi_k}T^j\ket{\psi_k}}^2 = \abs{n_0 + (-1)^{j}n_1}^2 = \delta_{0,j} + \delta_{1,j}\left(\frac{2\lambda_2 \cos k}{1+\lambda_2^2}\right)^2 \, ,
\end{equation}

and we see that there is only one moment that participates,

\begin{equation}
    \ln \tilde{Z}^{(1)}_{01}(0) = l  \int_0^\pi \frac{dk}{2\pi}\ln\left(\frac{2\lambda_2 \cos k}{1+\lambda_2^2}\right)^2 = l\ln\left(\frac{\lambda_2}{1+\lambda_2^2}\right) \, .
\end{equation}

Plugging this into the Frobenius distance of Eq.~\eqref{eq:frobqp} then gives the initial Frobenius distance of this state as

\begin{equation}
    \Delta F (\rho_2(\lambda_2),t=0) = \frac{1}{2} \left[ 1 - \left(\frac{\lambda_2}{1 + \lambda_2^2}\right)^{l} \,\right] ,
    \label{eq:psi2init}
\end{equation}

where the effect of superposition is exponentially suppressed with subsystem size $l$. Meanwhile, the Frobenius distance of the $\ket{\psi_4(\lambda_4)}$ states must be extracted from four-point correlators, which makes dependence of the initial Frobenius distance on $\lambda_4$ less straightforward. Instead, to prove that $\Delta F(\rho_{4}, t=0) > \Delta F(\rho_{2}, t=0)$, it will be sufficient for us to solve $\Delta F(\rho_4, t=0)$ for the choices $\lambda_4 = 0$ and $\lambda_4 \to \infty$, which correspond to the minimal and maximal symmetry breaking of the state. 

On the one hand, for $\lambda_4 = 0$, we have $\tilde{Z}^{(j)}(k) = \delta_{j\text{mod}2, 0}$, and plugging into Eq.~\eqref{eq:frobqp} yields $\Delta F(\rho_4(\lambda_4), t=0) = \sqrt{1/2}$. This coincides with the maximum value of $\Delta F(\rho_2(\lambda_2), t=0)$ found by setting either $\lambda_2 = 0$ or $\lambda_2 \to \infty$ in  Eq.~\eqref{eq:psi2init}. On the other hand, for $\lambda_4 = \infty$, we have $\tilde{Z}^{(j)}(k) = \delta_{j, 0}$ and $\Delta F(\rho_4(\lambda_4), t=0) = \sqrt{3/4}$. Since all other values of $\lambda_4$ correspond to intermediate symmetry breaking, we can therefore bound $\sqrt{1/2} \leq \Delta F(\rho_4(\lambda_4), t=0) \leq \sqrt{3/4}$ and assert that $\Delta F(\rho_2(\lambda_2), t=0)  \leq \Delta F(\rho_4(\lambda_4), t=0)$ for all $\lambda_2, \lambda_4$. 

\subsection{Long-time Limit}
\label{sm:longtime}

In the $\nu = 2$ case, the only non-zero moment given by the quasiparticle picture is

\begin{equation}
    \tilde{Z}^{(1)}_{01}(t) = \int^{\pi}_0 \frac{dk}{2\pi} \ln\left(\frac{2\lambda_2 \cos k}{1+\lambda_2^2}\right)^2 \max(l-2t\sin(k),0) \, .
\end{equation}

Expanding about $k^* = 0,\pi$ to leading order in $k$ and combining integrals, we find

\begin{equation}
    \tilde{Z}^{(1)}_{01}(t \gg 0) = \ln\left(\frac{2\lambda_2 } {1+\lambda_2^2}\right)^2 \int^{l/2t}_0 \frac{dk}{2\pi} \max(l-2tk,0) + O(k^2) = \ln\left(\frac{2\lambda_2}{1+\lambda_2^2}\right)^2 \cdot\frac{l^2}{8\pi t} + O(k^2) \, .
\end{equation}

This causes the Frobenius distance to decay at late times as

\begin{equation}
    \Delta F(\rho_2(\lambda_2), t \gg 0) = \left[\frac{1}{2} \left( 1 - \left( \frac{2\lambda_2}{1+\lambda_2^2}\right)^{l^2/8\pi t}\right)\right]^{1/2} \, ,
    \label{eq:psi2longtime}
\end{equation}

which correctly approaches $0$ from above as $t \to \infty$. Similarly, in the $\nu=4$ case we find

\begin{equation}
    \Delta F(\rho_4(\lambda_4), t \gg0) = \left[ \frac{1}{4} \left(3 - \Lambda_{1}^{l^2/8\pi t} - 2 \Lambda_{2}^{l^2/8\pi t}\right) \right]^{1/2}\, .
    \label{eq:psi4longtime}
\end{equation}

where we have introduced $\Lambda_j = f(j)^{1/4} g(j)^{1/\sqrt{2}}$, with $\tilde{Z}^{(j)}_{01}(\pi/4) = \tilde{Z}^{(j)}_{23}(\pi/4) = g(j) + O(k)$ and $\tilde{Z}^{(j)}_{02}(\pi/4) = \tilde{Z}^{(j)}_{13}(\pi/4) = f(j) + O(k)$. From Eqs.~\eqref{eq:psi2longtime} and \eqref{eq:psi4longtime}, the second condition for an asymmetry crossing, $\Delta F(\rho_{\nu_1}, t) < \Delta F(\rho_{\nu_2}, t)$ in the limit $t \to \infty$, can be written as

\begin{equation}
    \frac{2\lambda_2}{1 + \lambda_2^2} \leq \Lambda_2 \Lambda_1^2 \, .
    \label{eq:condlongtime}
\end{equation} 

Despite the apparent simplicity of this result, a general constraint between $\lambda_2$ and $\lambda_4$ would be too complicated to offer practical insight. Instead, we shall make the arbitrary choice to fix $\lambda_4 = 1$. This yields $f(1) = 5/9$, $f(2) = 1$, $g(1) = 21/25$ and $g(2) = 17/25$, resulting in the approximate inequalities $\lambda_2 < 0.234$ and $4.443 < \lambda_2$ for which Eq.~\eqref{eq:condlongtime} is satisfied, and hence for which an crossing of the Frobenius distance, according to the quasiparticle picture, is observed.

\end{document}